\documentclass[12pt]{article}
\usepackage{graphicx,cite,amsmath,epsfig}    
\usepackage{amssymb}
\usepackage{amsmath,color}
\usepackage{epsfig}
\usepackage{graphicx,epsf,epsfig}
\usepackage{footnote}
\usepackage{multirow}
\usepackage{enumitem}
\graphicspath{{figs/}}
\usepackage{float}
\usepackage{graphicx}
\usepackage{color}    
\usepackage{cite}
\usepackage{amsmath}
\usepackage{amssymb}
\usepackage{mathtools}

\def\beq{\begin{equation}}
\def\eeq{\end{equation}}
\def\nn{\nonumber}
\def\bea{\begin{eqnarray}}
\def\eea{\end{eqnarray}}
\def\ba{\begin{array}}                  
\def\ea{\end{array}}

\newcommand{\vt}{\vert}
\newcommand{\mc}{\mathcal}
\def\dg{\dagger}     

\include{paperdef}

\evensidemargin \oddsidemargin
\begin{document}

\thispagestyle{empty}

\def\thefootnote{\fnsymbol{footnote}}


\vspace{1cm}

\begin{center}
{\large\sc {\bf Non-Minimal Flavored ${\bf S}_{3}\otimes {\bf Z}_{2}$ Left-Right Symmetric Model}}

\vspace{0.4cm}

\vspace{1cm}

{
\sc Juan Carlos G\'omez-Izquierdo$^{1, 2, 3}$\footnote{email: jcgizquierdo1979@gmail.com}}

\vspace*{1cm}

{\sl $^{1}$
Tecnologico de Monterrey, Campus Estado de 
Mexico, Atizapan de Zaragoza, Estado de Mexico, Apartado Postal 52926, Mexico.\\
\sl $^{2}$ Instituto~de Ciencias Nucleares, Universidad~Nacional Aut\'onoma de M\'exico, 
M\'exico 3000, D.F., M\'exico.\\
\sl $^{3}$Instituto~de F{\'{\i}}sica, Universidad~Nacional Aut\'onoma de M\'exico, 
M\'exico 01000, D.F., M\'exico.\\}

\end{center}
\vspace*{0.2cm}
\begin{abstract}
We propose a non-minimal left-right symmetric model (LRSM) with Parity Symmetry where the fermion mixings arise as result of imposing an ${\bf S}_{3}\otimes {\bf Z}_{2}$ flavor symmetry, and an extra ${\bf Z}^{e}_{2}$ symmetry is considered to suppress some Yukawa couplings in the lepton sector. As a consequence, the effective neutrino mass matrix possesses approximately the $\mu-\tau$ symmetry. The breaking of the $\mu-\tau$ symmetry
induces sizable non zero $\theta_{13}$, and the deviation of $\theta_{23}$ from $45^{\circ}$ is strongly
controlled by an $\epsilon$ free parameter and the complex neutrino masses. Then, an analytic study on the extreme Majorana phases is done since these turn out to be relevant to enhance or suppress the reactor and atmospheric angle. So that we have constrained the parameter space for the $\epsilon$ parameter and the lightest neutrino mass that accommodate the mixing angles. The highlighted results are: a) the normal hierarchy is ruled out since the reactor angle comes out being tiny, for any values of the Majorana phases; b) for the inverted hierarchy there is one combination in the extreme phases where the values of the reactor and atmospheric angles are compatible up to $2, 3~\sigma$ of C. L., but the parameter space is tight; c) the model favors the degenerate ordering for one combination in the extreme Majorana phases. In this case, the reactor and atmospheric angle are compatible with the experimental data for a large set of values of the free parameters. Therefore, this model may be testable by the future result that the Nova and KamLAND-Zen collaborations will provide.

\end{abstract}

\def\thefootnote{\arabic{footnote}}
\setcounter{page}{0}
\setcounter{footnote}{0}

\newpage

\section{Introduction}
Currently, we know that neutrinos oscillate and have a tiny mass. In the theoretical framework of three active neutrinos, the difference of 
the squared neutrino masses for normal (inverted) hierarchy are given by~
$\Delta m^{2}_{21} \left( 10^{-5} \, \textrm{eV}^{2} \right) = 
7.60_{-0.18}^{+0.19},$ and $\left| \Delta m^{2}_{31} \right| \left( 10^{-3} \, \textrm{eV}^{2} \right)= 2.48_{-0.07}^{+0.05}~ (2.38_{-0.06}^{+0.05})$. Additionally, we have the values of the mixing angles $\sin^{2} \theta_{12} / 10^{-1} = 3.23 \pm 0.16,$  $\sin^{2} \theta_{23} / 10^{-1} = 5.67_{-1.24}^{+0.32}~(5.73_{-0.39}^{+0.25})$ and $\sin^{2} \theta_{13} / 10^{-2}= 2.26 \pm 0.12~(2.29 \pm 0.12)$~\cite{Forero:2014bxa}.At present, there is no yet solid evidence on the Dirac CP-violating phase and the ordering that respects the neutrino masses. The Nova \cite{Adamson:2017qqn} and KamLAND-Zen \cite{KamLAND-Zen:2016pfg} Collaborations can shed light on the hierarchy in the coming years.

In spite of the fact that the Standard Model (SM) works out almost perfectly, the neutrino experimental data can not be explained within this framework. If the neutrino sector opens the window to the new physics, then what is the new model and the extra ingredients that are needed to accommodate the masses and mixings ?. In this line of thought, a simplest route to include small neutrino masses and
mixings to the SM is to add the missing right-handed neutrinos (RHN's)
states  to the matter content, and then invoking the see-saw mechanism~\cite{Minkowski:1977sc, GellMann:1980vs, Yanagida:1979as, Mohapatra:1979ia, Schechter:1980gr, Mohapatra:1980yp, Schechter:1981cv}.
However, we should point out that the RHN mass scale is introduced by hand with no relation whatsoever to the Higgs mechanism that gives mass to all other fields. Nonetheless, this problem may be alleviated if the minimal extension of the SM is replaced by the left-right symmetric model (LRSM) \cite{Pati:1974yy, Mohapatra:1974gc, Senjanovic:1975rk, Senjanovic:1978ev, Mohapatra:1979ia} where the RHN's are already included in the matter content. Additionally, the see-saw mechanism comes in rather naturally in the context of left-right symmetric scenarios;
aside from other nice features, as for instance the recovery of Parity Symmetry, and the appearance of right-handed currents at high energy,
which also makes such extensions very appealing. Recently, the left-right scenarios have been revised~\cite{Chen:2013fna, Senjanovic:2014pva, Dev:2015kca, Chakrabortty:2016wkl, Dev:2016dja, Mitra:2016kov, Senjanovic:2016bya, Senjanovic:2015yea, Lindner:2016lpp, Patra:2015bga, Lindner:2016bgg} in order to make contact with the last experimental data of LHC. Moreover, the dark matter problem~\cite{Dev:2016xcp, Berlin:2016eem, Patra:2015vmp} and the diphoton excess anomaly~\cite{Hati:2016thk, Deppisch:2016scs, Dev:2015vjd, Dey:2015bur, Das:2015ysz} have been explored in this kind of scenarios.

Explaining the peculiar neutrino mixing pattern (besides the CKM mixing matrix) has been a hard task. Along this line, the mass textures have played an important role in trying to solve this puzzle~\cite{Fritzsch:1999ee}. In fact, discrete symmetries may be the missing ingredient to understand the mixings so that several groups have been proposed~\cite{Ishimori:2010au, Ishimori:2012zz, King:2013eh} to get in an elegant way the mass textures. In this line of thought, the ${\bf S}_{3}$ flavor symmetry, in particular, is a good candidate to handle the Yukawa couplings for leptons and quarks; and this has been studied exhaustively in different frameworks~\cite{Chen:2004rr, Felix:2006pn, Mondragon:2007af, Canales:2011ug, Canales:2012ix, Kubo:2012ty, Canales:2012dr, GonzalezCanales:2012kj, GonzalezCanales:2012za, Canales:2013ura, Canales:2013cga, Hernandez:2014lpa, Hernandez:2014vta, Hernandez:2015dga, Hernandez:2015zeh, Hernandez:2015hrt, Arbelaez:2016mhg, Hernandez:2013hea, CarcamoHernandez:2016pdu, Das:2014fea, Das:2015sca, Pramanick:2016mdp}. In most of these works, the meaning of the flavor has been extended to the scalar sector such that three Higgs doublets are required to accommodate the PMNS and CKM mixing matrices. 

Although there are too many flavored models in the literature, the LRSM has received few attention in the context of the flavored puzzle~\cite{GomezIzquierdo:2009id, Dev:2013oxa, Rodejohann:2015hka}. It is not an easy task to study the mixings in the LRSM since the structure of the gauge group increases the Yukawa sector parameters compared to the SM. However, as was shown in the early works, Parity Symmetry might reduce substantially the gauge and Yukawa couplings; this last issue gives the opportunity to calculate the right-handed CKM matrix \cite{Senjanovic:2016bya, Senjanovic:2015yea} which is crucial to study in great detail the $W_{R}$ gauge boson that comes out being a prediction of the LRMS. Then, it is fundamental to face the flavor puzzle in this kind of theoretical frameworks.

Therefore, we propose a non-minimal LRSM with Parity Symmetry where the fermion mixings arise as result of imposing an ${\bf S}_{3}\otimes {\bf Z}_{2}$ flavor symmetry, and an extra ${\bf Z}^{e}_{2}$ symmetry is considered to suppress some Yukawa couplings in the lepton sector. Additionally, a non conventional assignment is done for the matter content under the ${\bf S}_{3}$ symmetry and this is the clear difference between the previous studies and this one. As a consequence, in the lepton sector, the effective neutrino mass matrix possesses approximately the $\mu-\tau$ symmetry~\cite{Mohapatra:1998ka, Lam:2001fb, Kitabayashi:2002jd, Grimus:2003kq, Koide:2003rx, Fukuyama:1997ky, Gupta:2013it, Grimus:2012hu, Xing:2015fdg, Luo:2014upa, Ahn:2014gva, Rivera-Agudelo:2015vza, Zhao:2016orh, Biswas:2016yan}. The breaking of the $\mu-\tau$ symmetry
induces sizable non zero $\theta_{13}$, and the deviation of $\theta_{23}$ from $45^{\circ}$ is strongly
controlled by an $\epsilon$ free parameter and the complex neutrino masses.
Then, an analytic study on the extreme Majorana phases is done since these turn out to be relevant to enhance or suppress the reactor and atmospheric angle. Thus, we can constrain the parameter space for $\epsilon$ parameter and the lightest neutrino mass that accommodate the mixing angles. The highlighted results are: a) the normal hierarchy is ruled out since the reactor angle comes out being tiny, for any values of the Majorana phases; b) for the inverted hierarchy there is one combination in the extreme phases where the values of the reactor and atmospheric angles are compatible up to $2, 3~\sigma$ of C. L., but the parameter space is tight; c) the model favors the degenerate ordering for one combination in the extreme Majorana phases. In this case, the reactor and atmospheric angle are compatible with the experimental data for a large set of values of the free parameters. The quark sector will be discussed exhaustively in a future work, however, some preliminary results will be commented.

The paper is organized as follows: we present, in Sec. II,  the matter content of the model and also their respective assignment under the ${\bf S}_{3}$ symmetry. In addition, we briefly explain the scalar sector and argue about the need to include the ${\bf Z}^{e}_{2}$ symmetry. In Sec. III, the fermion mass matrices are obtained and we put attention on the lepton sector for getting the mixing matrices. We present, in Sec. IV, the PMNS matrix that the model predicts. Finally, we present an analytic study on the mixing angles  and our results  in Sec. V, and we  close our discussion  with a summary of conclusions.

\section{Flavored Left-Right Symmetric Model}
The minimal LRSM is based on the usual, $SU(3)_{c}\otimes SU(2)_{L}\otimes SU(2)_{R}\otimes U(1)_{B-L}$,  gauge symmetry where Parity Symmetry, $\mathcal{P}$, is assumed to be a symmetry a high energy but it is broken at electroweak scale since there are no right-handed currents. The matter fields and their respective quantum numbers (in parenthesis) under the gauge symmetry are given by
{\scriptsize
\begin{align}
Q_{(L, R)}&=\left(
\ba{c}
u \\
d \\
\ea
\right)_{(L, R)}\sim {\left(3, (2,1), (1,2), 1/3\right)},\quad
(L, R)=\left(
\ba{c}
\nu \\
\ell \\
\ea
\right)_{(L, R)}\sim { (1, (2,1), (1,2), -1)},\nonumber\\
\Phi&=\left(
\ba{cc}
\phi^{0} & \phi^{'+} \\
\phi^{-} & \phi^{'0} \\
\ea
\right)\sim \left(1, 2, 2, 0 
\right);\quad \Delta_{(L, R)}=\left(
\ba{cc}
\frac{\delta^{+}}{2} & \delta^{++} \\
\delta^{0} & -\frac{\delta^{+}}{2} \\
\ea
\right)_{(L, R)}\sim  \left(1, (3,1), (1,3), 2 \right). \label{eq1}
\end{align}}

The gauge invariant Yukawa mass term is given by
{\scriptsize
\begin{align}
-\mathcal{L}_{Y}=\bar{Q}_{L}\left[y^{q}\Phi+ \tilde{y}^{q}\tilde{\Phi} \right]Q_{R}+ \bar{L}\left[y^{\ell}\Phi+ \tilde{y}^{\ell}\tilde{\Phi} \right]R+ y^{L}\bar{L}\Delta_{L}L^{c}+y^{R}\bar{R}^{c}\Delta_{R}R+h.c. \label{yt}
\end{align}}
where the family indexes have been suppressed and $\tilde{\Phi}_{i}=-i\sigma_{2}\Phi^{\ast}_{i}i\sigma_{2}$. Here, Parity Symmetry will be assumed in the above Lagrangian, then, this requires that $\Psi_{i L}\leftrightarrow \Psi_{i R}$,
$\Phi_{i} \leftrightarrow \Phi^{\dg}_{i}$ and 
$\Delta_{i L}\leftrightarrow \Delta^{\dg}_{i R}$ for fermions and scalar fields, respectively. Thereby, the Yukawa couplings may reduce substantially and the gauge couplings too. In particular, for the former issue we have that $y=y^{\dagger}$, $\tilde{y}=\tilde{y}^{\dagger}$ and $y^{R}=y^{L}$. On the other hand, due to our purpose the scalar potential will be left aside. But, in the minimal LRMS the spontaneous symmetry breaking is as follows: Parity Symmetry is broken at the same scale where the $\Delta_{R}$ right-handed scale acquires its vacuum expectation value (vev). At the first stage, the RHN's are massive particles, then the rest of the particles turn out massive since the Higgs scalars get their vev's. Explicitly, 
 {\scriptsize
 	\begin{align}
 	\langle\Delta_{L, R}\rangle= \left(
 	\ba{cc}
 	0 & 0 \\
 	v_{L,R} & 0 \\
 	\ea
 	\right),\quad \langle\Phi\rangle= \left(
 	\ba{cc}
 	k & 0 \\
 	0 & k^{\prime}  \\
 	\ea
 	\right), \quad \langle \tilde{\Phi}\rangle= \left(
 	\ba{cc}
 	k^{\prime \ast} & 0 \\
 	0 & k^{\ast}  \\
 	\ea
 	\right).\label{eq6}
 	\end{align}}
As result, the Yukawa mass term is given by 
{\scriptsize
	\begin{align}
	-\mathcal{L}_{Y}=\bar{q}_{i L} \left({\bf M}_{q} \right)_{ij}q_{j R}+\bar{\ell}_{i L} \left( {\bf M}_{\ell}\right)_{ij}\ell_{j R}
	+\dfrac{1}{2}\bar{\nu}_{i L}\left({\bf M}_{\nu}\right)_{ij}\nu^{c}_{j L}+\dfrac{1}{2}\bar{\nu}^{c}_{i R}\left({\bf M}_{R}\right)_{ij}\nu_{j R}+h.c.\label{eq7}
	\end{align}}
where the type I see-saw mechanism has been realized, ${\bf M}_{\nu}=-{\bf M}_{D} {\bf M}^{-1}_{R} {\bf M}^{T}_{D}$; so that the ${\bf M}_{L}$ were neglected for simplicity.

In the present model, the Yukawa mass term will be controlled by the ${\bf S}_{3}$ flavor symmetry. The non-Abelian group ${\bf S}_{3}$ is the permutation group of three objects and this has three irreducible representations: two 1-dimensional, ${\bf 1}_{S}$ and ${\bf 1}_{A}$, and one 2-dimensional representation, ${\bf 2}$ (for a detailed study see \cite{Ishimori:2010au}). The multiplication rules among them are
{\scriptsize 
\begin{align}\label{rules}
&{\bf 1}_{S}\otimes {\bf 1}_{S}={\bf 1}_{S},\quad {\bf 1}_{S}\otimes {\bf 1}_{A}={\bf 1}_{A},\quad {\bf 1}_{A}\otimes {\bf 1}_{S}={\bf 1}_{A},\quad {\bf 1}_{A}\otimes {\bf 1}_{A}={\bf 1}_{A},\nn\\&
{\bf 1}_{S}\otimes {\bf 2}={\bf 2},\quad {\bf 1}_{A}\otimes {\bf 2}={\bf 2},\quad {\bf 2}\otimes {\bf 1}_{S}={\bf 2},\quad {\bf 2}\otimes {\bf 1}_{A}={\bf 2};\nn\\
&\begin{pmatrix}
a_{1} \\ 
a_{2}
\end{pmatrix}_{{\bf 2}}
\otimes
\begin{pmatrix}
b_{1} \\ 
b_{2}
\end{pmatrix}_{{\bf 2}} = 
\left(a_{1}b_{1}+a_{2}b_{2}\right)_{{\bf 1}_{S}} \oplus  \left(a_{1}b_{2}-a_{2}b_{1}\right)_{{\bf 1}_{A}} \oplus	
\begin{pmatrix}
a_{1}b_{2}+a_{2}b_{1} \\ 
a_{1}b_{1}-a_{2}b_{2}
\end{pmatrix}_{{\bf 2}}. 
\end{align}}
Having introduced briefly the gauge and the non-Abelian group, let us build the gauge and flavored Yukawa mass term. To do this, we will consider three Higgs bidoublets as well as three left-right triplets with the purpose of getting the mixing in the lepton sector. Here, we want to emphasize a clear difference between this model and the previous ones with the ${\bf S}_{3}$ symmetry. In our model, the quark and lepton families have been assigned in a different way under the irreducible representation of ${\bf S}_{3}$. Explicitly, for the former and the Higgs sector respectively, the first and second family have been put together in a flavor doublet ${\bf 2}$, and the third family is a singlet ${\bf 1}_{S}$. On the contrary, for the latter sector, the first family is a singlet ${\bf 1}_{S}$ and the second and third families are put in a doublet ${\bf 2}$. The advantage of making this choice is that the quark mass matrices may be put into two mass textures fashion that fit the CKM matrix very well; in the lepton sector, on the other hand, the appearance of the approximated $\mu-\tau$ symmetry in the effective neutrino mass matrix is good signal to understand the mixings. Remarkably, Nova collaboration  is testing the $\mu-\tau$ symmetry and some results have been released \cite{Adamson:2017qqn}.

The matter content of the model transforms in a not trivial way under the ${\bf S}_{3}$ symmetry and this is displayed in the table below. Here, the ${\bf Z_{2}}$ symmetry has been added in order to prohibit some Yukawa couplings in the lepton sector.
\begin{table}[ht]
\begin{center}
\begin{tabular}{|c|c|c|c|c|c|c|c|c|c|c|c|c|c|c|}
\hline \hline	
{\footnotesize Matter} & {\footnotesize $Q_{I (L, R)}$} & {\footnotesize $Q_{3 (L, R)}$} & {\footnotesize $(L_{1}, R_{1})$} & {\footnotesize $(L_{J}, R_{J})$}  & {\footnotesize $\Phi_{I}$} & {\footnotesize $\Phi_{3}$} &  {\footnotesize $\Delta_{I(L, R)}$ } & {\footnotesize $\Delta_{3 (L, R)}$}  \\ \hline
{\footnotesize \bf $S_{3}$} &  {\footnotesize \bf $2$} & {\footnotesize \bf $1_{S}$}   & {\footnotesize \bf $1_{S}$} & {\footnotesize \bf $2$} & {\footnotesize \bf $2$} & {\footnotesize \bf $1_{S}$} & {\footnotesize \bf $2$} & {\footnotesize \bf $1_{S}$} \\ \hline
{\footnotesize \bf $Z_{2}$} & {\footnotesize $1$} & {\footnotesize $1$}  &  {\footnotesize $1$} & {\footnotesize $-1$} & {\footnotesize $1$} & {\footnotesize $1$} & {\footnotesize \bf $-1$} & {\footnotesize \bf $1$} \\ \hline \hline
\end{tabular}\caption{Non-minimal left-right model. Here, $I=1,2$ and $J=2,3$.}
\end{center}
\end{table}
Thus, the most general Yukawa mass term, that respects the ${\bf S}_{3}\otimes {\bf Z}_{2}$ flavour symmetry and the gauge group, is given as
{\scriptsize
\begin{align}
-\mathcal{L}_{Y}&=y^{q}_{1}\left[\bar{Q}_{1 L}\left(\Phi_{1} Q_{2 R}+\Phi_{2} Q_{1 R}\right)+\bar{Q}_{2 L}\left(\Phi_{1} Q_{1 R}-\Phi_{2}Q_{2 R}\right)\right]+y^{q}_{2}\left[\bar{Q}_{1 L}\Phi_{3}Q_{1 R}+\bar{Q}_{2 L}\Phi_{3} Q_{2 R}\right]+y^{q}_{3}\left[\bar{Q}_{1 L}\Phi_{1}+\bar{Q}_{2 L}\Phi_{2}\right]Q_{3 R}\nn\\&+y^{q}_{4}\bar{Q}_{3 L}\left[\Phi_{1}Q_{1 R}+\Phi_{2}Q_{2 R}\right]+y^{q}_{5}\bar{Q}_{3 L}\Phi_{3}Q_{3 R}+\tilde{y}^{q}_{1}\left[\bar{Q}_{1 L}\left(\tilde{\Phi}_{1} Q_{2 R}+\tilde{\Phi}_{2}Q_{1 R}\right)+\bar{Q}_{2 L}\left(\tilde{\Phi}_{1} Q_{1 R}-\tilde{\Phi}_{2}Q_{2 R}\right)\right]\nn\\&+\tilde{y}^{q}_{2}\left[\bar{Q}_{1 L}\tilde{\Phi}_{3}Q_{1 R}+\bar{Q}_{2 L}\tilde{\Phi}_{3} Q_{2 R}\right]+\tilde{y}^{q}_{3}\left[\bar{Q}_{1 L}\tilde{\Phi}_{1}+\bar{Q}_{2 L}\tilde{\Phi}_{2}\right]Q_{3 R}+\tilde{y}^{q}_{4}\bar{Q}_{3 L}\left[\tilde{\Phi}_{1}Q_{1 R}+\tilde{\Phi}_{2}Q_{2 R}\right]+\tilde{y}^{q}_{5}\bar{Q}_{3 L}\tilde{\Phi}_{3}Q_{3 R}
+y^{\ell}_{1}\bar{L}_{1}\Phi_{3}R_{1}\nn\\&+y^{\ell}_{2}\left[(\bar{L}_{2}\Phi_{2}+\bar{L}_{3}\Phi_{1})R_{2}+(\bar{L}_{2}\Phi_{1}-\bar{L}_{3}\Phi_{2})R_{3} \right]+y^{\ell}_{3}\left[\bar{L}_{2}\Phi_{3}R_{2}+\bar{L}_{3}\Phi_{3}R_{3}\right]+
\tilde{y}^{\ell}_{1}\bar{L}_{1}\tilde{\Phi}_{3}R_{1}\nn\\&+\tilde{y}^{\ell}_{2}\left[(\bar{L}_{2}\tilde{\Phi}_{2}+\bar{L}_{3}\tilde{\Phi}_{1})R_{2}+(\bar{L}_{2}\tilde{\Phi}_{1}-\bar{L}_{3}\tilde{\Phi}_{2})R_{3} \right]+\tilde{y}^{ \ell}_{3}\left[\bar{L}_{2}\tilde{\Phi}_{3}R_{2}+\bar{L}_{3}\tilde{\Phi}_{3}R_{3}\right]
+y^{L}_{1}\bar{L}_{1}\Delta_{3  L}L^{c}_{1}+y^{L}_{2}\bar{L}_{1}\left[\Delta_{1 L }L^{c}_{2}+\Delta_{2 L}L^{c}_{3}\right]\nn\\&+y^{L}_{3} \left[ \bar{L}_{2}\Delta_{1 L}+\bar{L}_{3}\Delta_{2 L}\right]L^{c}_{1}+y^{L}_{4}\left[\bar{L}_{2}\Delta_{3 L}L^{c}_{2}+\bar{L}_{3}\Delta_{3 L}L^{c}_{3}\right]
+y^{R}_{1}\bar{R}^{c}_{1}\Delta_{3 R}R_{1}+ y^{R}_{2}\bar{R}^{c}_{1}\left[\Delta_{1 R}R_{2}+\Delta_{2 R}R_{3}\right]\nn\\&+y^{R}_{3}\left[\bar{R}^{c}_{2}\Delta_{1 R}+\bar{R}^{c}\Delta_{2 R}\right]R_{1}+ y^{R}_{4}\left[\bar{R}^{c}_{2}\Delta_{3 R}R_{2}+\bar{R}^{c}_{3}\Delta_{3 R}R_{3}\right]+h.c,\label{eq2}
\end{align}}
In this flavored model, we have to keep in mind that Parity Symmetry will be assumed in the above Lagrangian in such a way the number of Yukawa couplings is reduced. More even, we stress that
an extra symmetry ${\bf Z}^{e}_{2}$ is used to get a diagonal charged lepton and Dirac neutrino mass matrix whereas the Majorana mass matrices retain their forms. Explicitly, in the above Lagrangian, we demand that
{\footnotesize
\begin{align}
L_{3}\leftrightarrow-L_{3},\quad R_{3}\leftrightarrow-R_{3},\quad \Delta_{2 L}\leftrightarrow -\Delta_{2 L},\quad \Delta_{2 R}\leftrightarrow -\Delta_{2 R}.\label{exss}
\end{align}}
so that the terms $\bar{L}_{2}R_{3}$ and $\bar{L}_{3}R_{2}$ are absent in the lepton sector. As was already commented,
because of our interest in studying masses and mixings for fermions,
the scalar potential will not be analyzed for the moment. We ought to comment that this study is not trivial since the scalar sector has been augmented, so that the potential is rather complicated, but this study has to be done eventually since is crucial for theoretical and phenomenological purpose.

From Eq.(\ref{eq6}) and Eq.(\ref{eq2}), the mass matrices have the following structure
{\scriptsize
\begin{align}
{\bf M}_{q}=\begin{pmatrix}
a_{q}+b^{\prime}_{q} & b_{q} & c_{q} \\ 
b_{q} & a_{q}-b^{\prime}_{q} & c^{\prime}_{q} \\ 
f_{q} & f^{\prime}_{q} & g_{q}
\end{pmatrix},\quad {\bf M}_{\ell}=\begin{pmatrix}
a_{\ell} & 0 & 0 \\ 
0 & b_{\ell}+c_{\ell} & 0 \\ 
0 & 0 & b_{\ell}-c_{\ell}
\end{pmatrix}    ,\quad {\bf M}_{(L, R)}=\begin{pmatrix}
a_{(L, R)} & b_{(L, R)} & b^{\prime}_{(L, R)} \\ 
b_{(L, R)} & c_{(L, R)} & 0 \\ 
b^{\prime}_{(L, R)} & 0 & c_{(L, R)}
\end{pmatrix},\label{eq8} 
\end{align}}
where the $q= u, d$ and $\ell=e, \nu_{D}$. Explicitly, the matrix elements for quarks and leptons are given as
{\scriptsize
\begin{align}
&a_{u}=y^{q}_{2}k_{3}+\tilde{y}^{q}_{2}k^{\prime \ast}_{3},\quad b^{\prime}_{u}=y^{q}_{1}k_{2}+\tilde{y}^{q}_{1}k^{\prime \ast}_{2}  ,\quad b_{u}=y^{q}_{1}k_{1}+\tilde{y}^{q}_{1}k^{\prime \ast}_{1},\quad c_{u}=y^{q}_{3}k_{1}+\tilde{y}^{q}_{3}k^{\prime \ast}_{1},\,
c^{\prime}_{u}=y^{q}_{3}k_{2}+\tilde{y}^{q}_{3}k^{\prime \ast}_{2},\quad f_{u}= y^{\dg q}_{3}k_{1}+\tilde{y}^{\dg q}_{3}k^{\prime \ast}_{1},\nn\\ &f^{\prime}_{u}= y^{\dg q}_{3}k_{2}+\tilde{y}^{\dg q}_{3}k^{\prime \ast}_{2} ,\quad g_{u}=y^{q}_{5}k_{3}+\tilde{y}^{q}_{5}k^{\prime \ast}_{3},\quad
a_{d}=y^{q}_{2}k^{\prime}_{3}+\tilde{y}^{q}_{2}k^{\ast}_{3},\quad b^{\prime}_{d}=y^{q}_{1}k^{\prime}_{2}+\tilde{y}^{q}_{1}k^{\ast}_{2},\quad b_{d}=y^{q}_{1}k^{\prime}_{1}+\tilde{y}^{q}_{1}k^{\ast}_{1},\quad c_{d}=y^{q}_{3}k^{\prime}_{1}+\tilde{y}^{q}_{3}k^{\ast}_{1};\nn\\
&c^{\prime}_{d}=y^{q}_{3}k^{\prime}_{2}+\tilde{y}^{q}_{3}k^{\ast}_{2},\quad f_{d}= y^{\dg q}_{3}k^{\prime}_{1}+\tilde{y}^{\dg q}_{3}k^{\ast}_{1},\quad f^{\prime}_{d}= y^{\dg q}_{3}k^{\prime}_{2}+\tilde{y}^{\dg q}_{3}k^{\ast}_{2} ,\quad g_{u}=y^{q}_{5}k^{\prime}_{3}+\tilde{y}^{q}_{5}k^{\ast}_{3},\quad
a_{D}=y^{\ell}_{1}k_{3}+\tilde{y}^{\ell}_{1}k^{\prime \ast}_{3},\quad b_{D}=y^{\ell}_{3}k_{3}+\tilde{y}_{3}k^{\prime \ast}_{3},\nn\\ &c_{D}=y^{\ell}_{2}k_{2}+\tilde{y}_{2}k^{\prime \ast}_{2},\quad
a_{e}=y^{\ell}_{1}k^{\prime}_{3}+\tilde{y}^{\ell}_{1}k^{\ast}_{3},\quad b_{e}=y^{\ell}_{3}k^{\prime}_{3}+\tilde{y}_{3}k^{\ast}_{3},\quad c_{e}=y^{\ell}_{2}k^{\prime}_{2}+\tilde{y}_{2}k^{\ast}_{2},\quad
a_{(L, R)}=y^{R}_{1}v_{1(L, R)},\quad b_{(L, R)}=y^{R}_{2}v_{2(L, R)}\nn\\ &b^{\prime}_{(L, R)}=y^{R}_{2}v_{3(L, R)},\quad c_{(L, R)}=y^{R}_{4}v_{1(L, R)}.
\label{eq9}
\end{align}}
where Parity Symmetry has been considered. Remarkably, we will end up having a complex symmetric (diagonal) quark (lepton) mass matrix if the vev's are complex; in the literature this scenario is well known as {\bf pseudomanifest left-right symmetry} \cite{Langacker:1989xa, Harari:1983gq}. If the vev's are real, the quark (lepton) mass matrix is hermitian (real) and the number of CP phases are reduced, this framework is known as {\bf manifest left-right symmetry}\cite{Beg:1977ti, Langacker:1989xa} . In this work, we will discuss only the first framework and the second one will be studied in an extended version of the model and its consequences on the quark sector.

\section{Masses and Mixings}
In principle, in the mass matrices, we can reduce a further the number of free parameters considering certain alignment in the vev's, see Eq.(\ref{eq9}). So that, for the moment, we will assume that the vev's of $\Phi_{1}$ and $\Phi_{2}$ are degenerate. Explicitly, we demand that $k_{1}=k_{2}\equiv k$ and $k^{\prime}_{1}=k^{\prime}_{2}\equiv k^{\prime}$. Additionally, $v_{1 R}=v_{2 R}=v_{R}$. Therefore, we have:

{\bf Pseudomanisfest left-right theory}.
{\scriptsize
	\begin{align}
	{\bf M}_{q}=\begin{pmatrix}
	a_{q}+b_{q} & b_{q} & c_{q} \\ 
	b_{q} & a_{q}-b_{q} & c_{q} \\ 
	c_{q} & c_{q} & g_{q}
	\end{pmatrix},\quad {\bf M}_{\ell}=\begin{pmatrix}
	a_{\ell} & 0 & 0 \\ 
	0 & b_{\ell}+c_{\ell} & 0 \\ 
	0 & 0 & b_{\ell}-c_{\ell}
	\end{pmatrix}    ,\quad {\bf M}_{(L, R)}=\begin{pmatrix}
	a_{(L, R)} & b_{(L, R)} & b_{(L, R)} \\ 
	b_{(L, R)} & c_{(L, R)} & 0 \\ 
	b_{(L, R)} & 0 & c_{(L, R)}
	\end{pmatrix}. \label{eq10}
	\end{align}}

{\bf Manifest left-right theory}. 
{\scriptsize
	\begin{align}
	{\bf M}_{q}=\begin{pmatrix}
	a_{q}+b_{q} & b_{q} & c_{q} \\ 
	b_{q} & a_{q}-b_{q} & c_{q} \\ 
	c^{\ast}_{q} & c^{\ast}_{q} & g_{q}
	\end{pmatrix},\quad {\bf M}_{\ell}=\begin{pmatrix}
	a_{\ell} & 0 & 0 \\ 
	0 & b_{\ell}+c_{\ell} & 0 \\ 
	0 & 0 & b_{\ell}-c_{\ell}
	\end{pmatrix}    ,\quad {\bf M}_{(L, R)}=\begin{pmatrix}
	a_{(L, R)} & b_{(L, R)} & b_{(L, R)} \\ 
	b_{(L, R)} & c_{(L, R)} & 0 \\ 
	b_{(L, R)} & 0 & c_{(L, R)}
	\end{pmatrix}. \label{eq10.m}
	\end{align}}
As was already commented the full analysis of the quark masses and mixings will be left aside for this moment. However, we just make some comments. In the {\bf pseudomanifest} framework, the ${\bf M}_{q}$ mass matrix may be put into two mass textures fashion that fit the CKM matrix very well. In similar way, the {\bf manifest} framework is tackled. For this case, the quark mixing matrix has fewer free parameters than the above framework since this is hermitian; the study, and its predictions on the mixing angles is work in progress.

\subsection{Charged Leptons}
The ${\bf M}_{e}$ mass matrix is complex and diagonal then one could identify straight the physical masses, however, we will make a similarity transformation in order to prohibit a fine tuning in the free parameters. What we mean is the following, the ${\bf M}_{e}$ mass matrix is diagonalized by ${\bf U}_{e L}={\bf S}_{23}{\bf P}_{e}$ and ${\bf U}_{e R}={\bf S}_{23}{\bf P}^{\dg}_{e}$, this is, ${\hat{\bf M}_{e}}=\textrm{diag.}(\vt m_{e}\vt, \vt m_{\mu}\vt,\vt m_{\tau}\vt)={\bf U}^{\dg}_{e L}{\bf M}_{e}{\bf U}_{e R} ={\bf P}^{\dg}{\bf m}_{e}{\bf P}^{\dg}_{e}$ with ${\bf m}_{e}={\bf S}^{T}_{23}{\bf M}_{e}{\bf S}_{23}$. After factorizing the phases, we have ${\bf m}_{e}={\bf P}_{e}{\bf \bar{m}_{e}}{\bf P}_{e}$~ where
 {\scriptsize
 	\begin{align}
 	{\bf m}_{e}=\textrm{diag.}(m_{e}, m_{\mu}, m_{\tau}),\quad {\bf S}_{23}=\begin{pmatrix}
 	1 & 0 & 0 \\ 
 	0 & 0 & 1 \\ 
 	0 & 1 & 0
 	\end{pmatrix},\quad {\bf P}_{e}=\textrm{diag.}(e^{i\eta_{e}}, e^{i\eta_{\mu}}, e^{i\eta_{\tau}}) \label{eue}
 	\end{align}	}
As result, one obtains that $\vt m_{e}\vt=\vt a_{e}\vt$, $\vt m_{\mu}\vt=\vt b_{e}-c_{e}\vt$ and $\vt m_{\tau}\vt=\vt b_{e}+c_{e}\vt$. 

\subsection{Neutrinos}
On the other hand, the ${\bf M}_{\nu}$ effective neutrino mass matrix is given as
{\scriptsize
\begin{align}
{\bf M}_{\nu}=\begin{pmatrix}
\mc{X}a^{2}_{D}& -a_{D}\mc{Y}(b_{D}+c_{D})  & -a_{D}\mc{Y}(b_{D}-c_{D}) \\ 
-a_{D}\mc{Y}(b_{D}+c_{D})& \mc{W}(b_{D}+c_{D})^{2}  & \mc{Z}(b^{2}_{D}-c^{2}_{D})  \\ 
-a_{D}\mc{Y}(b_{D}-c_{D})& \mc{Z}(b^{2}_{D}-c^{2}_{D})  & \mc{W}(b_{D}-c_{D})^{2}
\end{pmatrix}\quad\textrm{where}\quad {\bf M}^{-1}_{R}\equiv\begin{pmatrix}
\mc{X}& -\mc{Y} & -\mc{Y} \\ 
-\mc{Y} & \mc{W} & \mc{Z} \\ 
-\mc{Y} & \mc{Z} & \mc{W} \label{efm}
\end{pmatrix} 
\end{align}}
Now as hypothesis, we will assume that $b_{D}$ is larger than $c_{D}$, in this way the effective mass matrix can be written as
{\scriptsize
\begin{align}
{\bf M}_{\nu}\equiv\begin{pmatrix}
A_{\nu}& -B_{\nu}(1+\epsilon) & -B_{\nu}(1-\epsilon) \\ 
-B_{\nu}(1+\epsilon)& C_{\nu}(1+\epsilon)^{2} & D_{\nu}(1-\epsilon^{2}) \\ 
-B_{\nu}(1-\epsilon)& D_{\nu}(1-\epsilon^{2})  & C_{\nu}(1-\epsilon)^{2}\label{efm2}
\end{pmatrix} 
\end{align}}
where $A_{\nu}\equiv \mc{X}a^{2}_{D}$, $B_{\nu} \equiv\mc{Y}a_{D}b_{D}$, $C_{\nu}\equiv\mc{W}b^{2}_{D}$ and $D_{\nu}\equiv\mc{Z}b^{2}_{D}$ are complex. Besides, $\epsilon\equiv c_{D}/b_{D}$ is complex too. Here, we want to stress that the last parameter will be considered as a perturbation to the effective mass matrix such that $\vert \epsilon \vert \lll 1$. To be more specific, $\vert \epsilon \vert\leq 0.3$ in order to break softly the $\mu-\tau$  symmetry. So that, hereafter, we will neglect the $\epsilon^{2}$ quadratic terms in the above matrix. Having done this, we go back to the effective neutrino mass matrix. In order to cancel the ${\bf S}_{23}$ contribution that comes from the charged lepton sector, we make the following to ${\bf M}_{\nu}$. We know that $\hat{\bf M}_{\nu}=\textrm{diag.}(m_{\nu_{1}}, m_{\nu_{2}}, m_{\nu_{3}})={\bf U}^{\dg}_{\nu}{\bf M}_{\nu}{\bf U}^{\ast}_{\nu}$, then ${\bf U}_{\nu}={\bf S}_{23}{\bf \mc{U}_{\nu}}$ where the latter mixing matrix will be obtained below. Then, $\hat{\bf M}_{\nu}={\bf \mc{U}^{\dg}_{\nu}}{\bf \mc{M}_{\nu}}{\bf \mc{U}^{\ast}_{\nu}}$ with
{\scriptsize
	\begin{align}
	{\bf \mc{M}_{\nu}}= {\bf S}^{T}_{23}{\bf M}_{\nu}{\bf S}_{23}\approx\begin{pmatrix}
	A_{\nu}& -B_{\nu}(1-\epsilon) & -B_{\nu}(1+\epsilon) \\ 
	-B_{\nu}(1-\epsilon)& C_{\nu}(1-2\epsilon) & D_{\nu} \\ 
	-B_{\nu}(1+\epsilon)& D_{\nu} & C_{\nu}(1+2\epsilon)\label{efm3}
	\end{pmatrix}
	\end{align}	}
When the $\epsilon$ parameter is switched off the effective mass matrix, which is denoted by ${\bf \mc{M}^{0}_{\nu}}$, possesses the $\mu-\tau$ symmetry and this is diagonalized by
{\scriptsize
	\begin{align}
	{\bf \mc{U}}^{0}_{\nu}=\begin{pmatrix}
	\cos{\theta}_{\nu}~e^{i(\eta_{\nu}+\pi)} & \sin{\theta}_{\nu}~e^{i(\eta_{\nu}+\pi)}  & 0 \\ 
	-\frac{\sin{\theta}_{\nu}}{\sqrt{2}}& \frac{\cos{\theta}_{\nu}}{\sqrt{2}} & -\frac{1}{\sqrt{2}} \\ 
	-\frac{\sin{\theta}_{\nu}}{\sqrt{2}}& \frac{\cos{\theta}_{\nu}}{\sqrt{2}} & \frac{1}{\sqrt{2}}
	\end{pmatrix} \label{ubm}
	\end{align}	}
where the ${\bf \mc{M}^{0}_{\nu}}$ matrix elements are fixed in terms of the complex neutrinos physical masses, the $\theta_{\nu}$ free parameter and the $\eta_{\nu}$ Dirac CP phase. To be more explicit,
{\scriptsize
\begin{align}
A_{\nu}&=(m^{0}_{\nu_{1}}\cos^{2}{\theta}_{\nu}+m^{0}_{\nu_{2}}\sin^{2}{\theta}_{\nu})e^{2i(\eta_{\nu}+\pi)},\quad -B_{\nu}=\frac{\sin{2\theta_{\nu}}}{\sqrt{8}}(m^{0}_{\nu_{2}}-m^{0}_{\nu_{1}})e^{i(\eta_{\nu}+\pi)};\nn\\ C_{\nu}&=\frac{1}{2}(m^{0}_{\nu_{1}}\sin^{2}{\theta}_{\nu}+m^{0}_{\nu_{2}}\cos^{2}{\theta}_{\nu}+m^{0}_{\nu_{3}}),\quad
D_{\nu}=\frac{1}{2}(m^{0}_{\nu_{1}}\sin^{2}{\theta}_{\nu}+m^{0}_{\nu_{2}}\cos^{2}{\theta}_{\nu}-m^{0}_{\nu_{3}}).\label{mne}
\end{align}	}
Including the $\epsilon$ parameter we can write the effective mass matrix as ${\bf \mc{M}}_{\nu}={\bf \mc{M}^{0}_{\nu}}+{\bf \mc{M}^{\epsilon}_{\nu}}$ where the second matrix contains the perturbation, then, when we apply ${\bf\mc{U}^{0}_{\nu}}$ one gets ${\bf \mc{M}}_{\nu}={\bf \mc{U}^{0\dg}_{\nu}}({\bf \mc{M}^{0}_{\nu}}+{\bf \mc{M}^{\epsilon}_{\nu}}){\bf \mc{U}^{0 \ast}_{\nu}}$. Explicitly
{\scriptsize
\begin{align}
{\bf \mc{M}}_{\nu}=\textrm{Diag.}(m^{0}_{\nu_{1}}, m^{0}_{\nu_{2}}, m^{0}_{\nu_{3}})+\begin{pmatrix}
0 &  &-\sin{\theta_{\nu}}(m^{0}_{\nu_{3}}+m^{0}_{\nu_{1}})\epsilon \\ 
0 & 0 & \cos{\theta_{\nu}}(m^{0}_{\nu_{3}}+m^{0}_{\nu_{2}})\epsilon \\ 
-\sin{\theta_{\nu}}(m^{0}_{\nu_{3}}+m^{0}_{\nu_{1}})\epsilon& \cos{\theta_{\nu}}(m^{0}_{\nu_{3}}+m^{0}_{\nu_{2}})\epsilon & 0\label{mnp}
\end{pmatrix} 
\end{align}	}
The contribution of second matrix to the mixing one is given by
{\scriptsize
\begin{align}
{\bf \mc{U}}^{\epsilon}_{\nu}\approx\begin{pmatrix}
N_{1}&  0 & -N_{3}\sin{\theta}r_{1}\epsilon \\ 
0 & N_{2} & N_{3}\cos{\theta_{\nu}}r_{2}\epsilon \\ 
N_{1}\sin{\theta_{\nu}}r_{1}\epsilon & -N_{2}\cos{\theta_{\nu}}r_{2}\epsilon & N_{3}
\end{pmatrix} \label{ubpr}
\end{align}}
where we have defined the complex mass ratios $r_{(1, 2)}\equiv (m^{0}_{\nu_{3}}+m^{0}_{\nu_{(1, 2)}})/(m^{0}_{\nu_{3}}-m^{0}_{\nu_{(1, 2)}})$. Here, $N_{1}$, $N_{2}$ and $N_{3}$ are the normalization factors which are given as
{\scriptsize
\begin{align}
N_{1}=\left(1+\sin^{2}{\theta_{\nu}}\vert r_{1}\epsilon \vert^{2}\right )^{-1/2},\quad N_{2}=\left(1+\cos^{2}{\theta_{\nu}}\vert r_{2}\epsilon \vert^{2}\right)^{-1/2},\quad N_{3}=\left(1+\sin^{2}{\theta_{\nu}}\vert r_{1}\epsilon \vert^{2}+\cos^{2}{\theta_{\nu}}\vert r_{2}\epsilon \vert^{2}\right)^{-1/2}.\label{nofac}
\end{align}}
Finally, the effective mass matrix given in Eq.(\ref{efm2}) is diagonalized approximately by ${\bf U}_{\nu}\approx {\bf S}_{23}{\bf \mc{U}^{0}_{\nu}{\bf \mc{U}^{\epsilon}_{\nu}}}$. Therefore, the theoretical PMNS mixing matrix is written as $V_{PMNS}={\bf U}^{\dg}_{e L}{\bf U}_{\nu}\approx {\bf P}^{\dg}_{e}{\bf \mc{U}^{0}_{\nu}{\bf \mc{U}^{\epsilon}_{\nu}}}$. 

\section{PMNS Mixing Matrix}
The PMNS mixing matrix is given explicitly as
{\scriptsize
\begin{align}
{\bf V}_{PMNS}={\bf P}^{\dagger}_{e}\begin{pmatrix}
\cos{\theta_{\nu}}N_{1}e^{i(\eta_{\nu}+\pi)}& \sin{\theta_{\nu}}N_{2}e^{i(\eta_{\nu}+\pi)} & \sin{2\theta_{\nu}}\frac{N_{3}}{2}(r_{2}-r_{1})\epsilon e^{i(\eta_{\nu}+\pi)} \\ 
-\frac{\sin{\theta_{\nu}}}{\sqrt{2}}N_{1}(1+r_{1}\epsilon)& \frac{\cos{\theta_{\nu}}}{\sqrt{2}}N_{2}(1+r_{2}\epsilon) & -\frac{N_{3}}{\sqrt{2}}\left[1-\epsilon~r_{3}\right] \\ 
-\frac{\sin{\theta_{\nu}}}{\sqrt{2}}N_{1}(1-r_{1}\epsilon)& \frac{\cos{\theta_{\nu}}}{\sqrt{2}}N_{2}(1-r_{2}\epsilon)  & \frac{N_{3}}{\sqrt{2}}\left[1+\epsilon~ r_{3}\right]
\end{pmatrix} \label{pmma}
\end{align}	}
where $r_{3}\equiv r_{2}\cos^{2}{\theta_{\nu}}+r_{1}\sin^{2}{\theta_{\nu}}$. On the other hand,
comparing the magnitude of entries  ${\bf V}_{PMNS}$ with the mixing matrix in the 
standard parametrization of the PMNS, we obtain the following expressions for 
the lepton mixing angles
{\scriptsize
\begin{align}
\sin^{2}{\theta}_{13}&=\vert {\bf V}_{13}\vert^{2} =\frac{\sin^{2}{2\theta_{\nu}}}{4}N^{2}_{3}\vert \epsilon \vert^{2}~\vert r_{2}-r_{1} \vert^{2},\quad
\sin^{2}{\theta}_{23}=\dfrac{\vert {\bf V}_{23}\vert^{2}}{1-\vert {\bf V}_{13}\vert^{2}}=\dfrac{N^{2}_{3}}{2}\frac{\vert 1-\epsilon~r_{3} \vert ^{2}}{1- \sin^{2}{\theta_{13}}},\nn\\
\sin^{2}{\theta_{12}}&=\dfrac{\vert {\bf V}_{12}\vert^{2}}{1-\vert {\bf V}_{13}\vert^{2}}=
\dfrac{N^{2}_{2}\sin^{2}{\theta_{\nu}}}{1-\sin^{2}{\theta}_{13}}.\label{mixang}
\end{align} }
As can be noticed, if $\epsilon$ vanishes, one would recover the exact $\mu-\tau$ symmetry where $\theta_{12}=0^{\circ}$ and $\theta_{23}=45^{\circ}$. Additionally, we have to point out that the reactor and atmospheric angles depend strongly on the neutrino masses ratios so that these angles are sensitive to the Majorana phases. At the same time, the reactor angle does not depend on the phase of the parameter $\epsilon$, but on the other hand, the atmospheric one has a clear dependency on this phase.

\section{Analytic Study and Results}
In order to make an analytic study on the above formulas, let us emphasize that we are working in a perturbative regime which means that~$\vert \epsilon \vert \leq 0.3$. Then $N_{i}$ normalization factors should be the order of $1$ so that, as is usual in models where the $\mu-\tau$ symmetry is broken softly, the solar angle is directly related to the free parameter $\theta_{\nu}$, as can be seen in Eq. (\ref{mixang}). Therefore, at the leading order we have that
\begin{align}
\sin^{2}{\theta_{12}}=\sin^{2}{\theta_{\nu}},\qquad\textrm{then},\qquad \theta_{12}=\theta_{\nu}.\label{sol}
\end{align}
Therefore, along the analytic study we will consider that $\sin{\theta_{\nu}}\approx 1/\sqrt{3}$ which is a good approximation to the solar angle. Additionally, we will analyze the extreme Majorana phases for the complex neutrino masses for each hierarchy. What we mean by extreme phases is that these can be either $0$ or $\pi$. Explicitly, $m^{0}_{\nu_{i}}=\pm \vert m^{0}_{\nu_{i}}\vert $, for $i=1, 2, 3$, where $\vert m^{0}_{\nu_{i}}\vert$ is the absolute mass. As we will see, these phases can be relevant to enhance  or suppress the reactor and atmospheric angles. In the following, the lightest neutrino mass and the $\vert \epsilon \vert$ parameter will be constrained.

{\bf Normal hierarchy}. From experimental data, the absolute neutrino masses are $\vert m^{0}_{\nu_{3}}\vert= \sqrt{\Delta m^{2}_{31}+\vert m^{0 }_{\nu_{1}}\vert^{2}}$ and $\vert m^{0}_{\nu_{2}}\vert=\sqrt{\Delta m^{2}_{21}+\vert m^{0}_{\nu_{1}}\vert^{2} }$. Now, the mass ratios $r_{1}$, $r_{2}$ and $r_{3}$ can be approximated as follows
{\scriptsize
	\begin{align}
	r_{1}&\approx 1+2\frac{m^{0}_{\nu_{1}}}{m^{0}_{\nu_{3}}}\approx 1,\qquad r_{2}\approx 1+2\frac{m^{0}_{\nu_{2}}}{m^{0}_{\nu_{3}}},\qquad r_{3}\approx 1+2\frac{m^{0}_{\nu_{2}}}{m^{0}_{\nu_{3}}}\cos^{2}{\theta_{\nu}}\label{massrat}
	\end{align}}
as results of this, we obtain
{\scriptsize
	\begin{align}
	\sin^{2}{\theta}_{13}&\approx\sin^{2}{2\theta_{\nu}}\vt \epsilon \vt^{2} \left|\frac{m^{0}_{\nu_{2}}}{m^{0}_{\nu_{3}}}\right|^{2},\quad
	\sin^{2}{\theta}_{23}\approx \dfrac{1}{2}\dfrac{ \left| 1-\epsilon \left(1+2\frac{m^{0}_{\nu_{2}}}{m^{0}_{\nu_{3}}}\cos^{2}{\theta_{\nu}}
		\right) \right|^{2} }{1-\sin^{2}{\theta}_{13}}.\label{annh}
	\end{align}}
As can be noticed, if the strict normal hierarchy is assumed then the reactor angle comes out being very small since $\vert m^{0}_{\nu_{2}}/m^{0}_{\nu_{3}}\vert^{2}\approx \Delta m^{2}_{21}/\Delta m^{2}_{31}$, and $\vt \epsilon \vt \leq 0.3$. This holds for any extreme Majorana phases in the neutrino masses and this result does not change substantially if the $m^{0}_{\nu_{1}}$ is non-zero. Therefore, the normal spectrum is ruled out for~$\vert \epsilon \vert \leq 0.3$.

{\bf Inverted hierarchy}. In this case, we have that $\vert m^{0}_{\nu_{2}}\vert=\sqrt{\Delta m^{2}_{13}+\Delta m^{2}_{21}+\vert m^{0 }_{\nu_{3}}\vert^{2}}$ and $\vert m^{0}_{\nu_{1}}\vert=\sqrt{\Delta m^{2}_{13}+\vert m^{0}_{\nu_{3}}\vert^{2}}$. The mass ratios $r_{1}$, $r_{2}$ and $r_{3}$ are written approximately as
{\scriptsize
\begin{align}
r_{(1, 2)}&\approx -\left(1+2\frac{m^{0}_{\nu_{3}}}{m^{0}_{\nu_{(1,2)}}}\right),\quad r_{2}-r_{1}\approx 2m^{0}_{\nu_{3}}\left[\frac{m^{0}_{\nu_{2}}-m^{0}_{\nu_{1}}}{m^{0}_{\nu_{2}}m^{0}_{\nu_{1}}}\right],\quad r_{3}\approx -\left[1+2\frac{m^{0}_{\nu_{3}}\left(m^{0}_{\nu_{1}}\cos^{2}{\theta_{\nu}}+m^{0}_{\nu_{2}}\sin^{2}{\theta_{\nu}}\right) }{m^{0}_{\nu_{2}}m^{0}_{\nu_{1}}}\right]\label{massratin}
\end{align}}
Due to the  mass difference $m^{0}_{\nu_{2}}-m^{0}_{\nu_{1}}$ in the factor $r_{2}-r_{1}$, the reactor angle can be small or large since the relative signs in these two masses may conspire to achieve it. Then, there are four independent cases where the signs in the masses can affect substantially the mixing angles:
 \begin{itemize}
 	\item {\bf Case A}. If $m_{\nu_{i}}> 0$.
 	{\scriptsize
 		\begin{align}
 		r_{2}-r_{1}\approx 2\vert m^{0}_{\nu_{3}}\vert \left[\frac{\vert m^{0}_{\nu_{2}}\vert -\vert m^{0}_{\nu_{1}}\vert }{\vert m^{0}_{\nu_{2}}\vert \vert m^{0}_{\nu_{1}}\vert}\right],\quad r_{3}\approx -\left[1+2\frac{\vert m^{0}_{\nu_{3}}\vert \left(\vert m^{0}_{\nu_{2}}\vert \sin^{2}{\theta_{\nu}}+\vert m^{0}_{\nu_{1}}\vert \cos^{2}{\theta_{\nu}}\right) }{\vert m^{0}_{\nu_{2}}\vert \vert m^{0}_{\nu_{1}}\vert} \right]\label{caih}
 		\end{align}}
 	\item {\bf Case B}. If $m^{0}_{\nu_{(2, 1)}}> 0$ and $m^{0}_{\nu_{3}}<0$.
 	{\scriptsize
 		\begin{align}
 		r_{2}-r_{1}\approx -2\vert m^{0}_{\nu_{3}}\vert \left[\frac{\vert m^{0}_{\nu_{2}}\vert -\vert m^{0}_{\nu_{1}}\vert }{\vert m^{0}_{\nu_{2}}\vert \vert m^{0}_{\nu_{1}}\vert}\right],\quad r_{3}\approx -\left[1-2\frac{\vert m^{0}_{\nu_{3}}\vert \left(\vert m^{0}_{\nu_{2}}\vert \sin^{2}{\theta_{\nu}}+\vert m^{0}_{\nu_{1}}\vert \cos^{2}{\theta_{\nu}}\right) }{\vert m^{0}_{\nu_{2}}\vert \vert m^{0}_{\nu_{1}}\vert} \right]\label{cbih}
 		\end{align}}
 	\item {\bf Case C}. If If $m^{0}_{\nu_{(3, 2)}}> 0$ and $m^{0}_{\nu_{1}}<0$. 
 	{\scriptsize
 		\begin{align}
 		r_{2}-r_{1}\approx -2\vert m^{0}_{\nu_{3}}\vert \left[\frac{\vert m^{0}_{\nu_{2}}\vert +\vert m^{0}_{\nu_{1}}\vert }{\vert m^{0}_{\nu_{2}}\vert \vert m^{0}_{\nu_{1}}\vert}\right],\quad r_{3}\approx -\left[1-2\frac{\vert m^{0}_{\nu_{3}}\vert \left(\vert m^{0}_{\nu_{2}}\vert \sin^{2}{\theta_{\nu}}-\vert m^{0}_{\nu_{1}}\vert \cos^{2}{\theta_{\nu}}\right)}{\vert m^{0}_{\nu_{2}}\vert \vert m^{0}_{\nu_{1}}\vert} \right]\label{ccih}
 		\end{align}} 
 	\item {\bf Case D}. If $m^{0}_{\nu_{2}}>0$ and $m^{0}_{\nu_{(3, 1)}}< 0$.
 	{\scriptsize
 		\begin{align}
 		r_{2}-r_{1}\approx 2\vert m^{0}_{\nu_{3}}\vert \left[\frac{\vert m^{0}_{\nu_{2}}\vert +\vert m^{0}_{\nu_{1}}\vert }{\vert m^{0}_{\nu_{2}}\vert \vert m^{0}_{\nu_{1}}\vert}\right],\quad r_{3}\approx -\left[1+2\frac{\vert m^{0}_{\nu_{3}}\vert \left(\vert m^{0}_{\nu_{2}}\vert \sin^{2}{\theta_{\nu}}-\vert m^{0}_{\nu_{1}}\vert \cos^{2}{\theta_{\nu}}\right)}{\vert m^{0}_{\nu_{2}}\vert \vert m^{0}_{\nu_{1}}\vert} \right]\label{cdih}
 		\end{align}}
 \end{itemize}
Noticing, if the strict inverted hierarchy were realized, $r_{2}-r_{1}=0$ and $r_{3}=-1$, we would have that $\sin^{2}{\theta_{13}}=0$ and $\sin^{2}{\theta_{23}}=N^{2}_{3}\vert 1+\epsilon \vert^{2}/2$, which is not compatible with the observations. Nonetheless, this strict ordering allows us to infer that the $\vert \epsilon \vert e^{i\alpha_{\epsilon}} $ parameter magnitude has to be small in order to deviate sufficiently the atmospheric angle from $45^{\circ}$, and the same time, this has to be enough large to enhance the reactor one. In here, the $\alpha_{\epsilon}$ associated phase determines if we are above or below of $45^{\circ}$. Along this line, Nova experiment has discarded the lower octant \cite{Adamson:2017qqn}. On the contrary, if the constraint, on the lightest neutrino mass, is relaxed, the reactor angle comes out being non zero and the atmospheric one has an extra contribution, $r_{3}$, which can enlarge or reduce the $\vert \epsilon \vert$ magnitude since this may be greater or minor than $1$. So that, the factor $\vert \epsilon~r_{3}\vert $ might deviate drastically the atmospheric angle beyond of $45^{\circ}$. 

Notice that, roughly speaking, the reactor angle turns out being equal for the {\bf Cases A} and {\bf B}, and also, for {\bf C} and {\bf D}. The key difference among them comes from the atmospheric angle as can be seen in Eq.(\ref{caih}-\ref{cdih}). Now, from the absolute value of the neutrino masses we have $\vert m^{0}_{\nu_{2}}\vert \approx \vert m^{0}_{\nu_{1}}\vert (1+2R_{1})$, then
 {\scriptsize
 	\begin{align}
 	\vert m^{0}_{\nu_{2}}\vert -\vert m^{0}_{\nu_{1}}\vert \approx 2\vert m^{0}_{\nu_{1}}\vert R_{1} ,\quad \vert m^{0}_{\nu_{2}}\vert +\vert m^{0}_{\nu_{1}}\vert \approx 2 \vert m^{0}_{\nu_{1}}\vert\left[1+R_{1}\right],\quad \vert m^{0}_{\nu_{1}}\vert \vert m^{0}_{\nu_{2}}\vert\approx \vert m^{0}_{\nu_{1}}\vert^{2}\left[1+2 R_{1} \right],\label{masdih}
 	\end{align}}
 where, $R_{1}\equiv \Delta m^{2}_{21}/4 \vert m^{0}_{\nu_{1}}\vert^{2}\approx \mathcal{O}(10^{-3})$, if the $\vert m^{0}_{\nu_{3}}\vert$ lightest neutrino mass is tiny.
 Therefore, for the {\bf Cases A} and {\bf B}, we have 
{\scriptsize
\begin{align}
\sin^{2}{\theta_{13}}\approx \frac{32}{9} | \epsilon |^{2} R^{2}_{1} \left| \frac{m^{0}_{\nu_{3}}}{m^{0}_{\nu_{1}}} \right|^{2},\qquad \sin^{2}{\theta_{23}}\approx \frac{1}{2}\frac{\left|1+\epsilon \left(1\pm 2 \left|\frac{m^{0}_{\nu_{3}}}{m^{0}_{\nu_{1}}}\right|\right)\right|^{2}}{1-\sin^{2}{\theta_{13}}}\label{reica}
\end{align}	} 
where the upper (lower) sign, in the atmospheric angle, stands for the {\bf Case A} ({\bf Case B}). Here, we have to keep in mind that $\vert m^{0}_{\nu_{3}}\vert/\vert m^{0}_{\nu_{1}}\vert< 1$ so that we can conclude that the first two scenarios are ruled out since that the reactor angle is proportional to the small quantity $(\vert m^{0}_{\nu_{3}}\vert/\vert m^{0}_{\nu_{1}}\vert)R_{1}\vert \epsilon \vert $, where $\vert \epsilon \vert \leq 0.3$. 
 
For the {\bf Case C} ( {\bf Case D}) the corresponding sign is the upper (lower), then the mixing angles are given as
 {\scriptsize
\begin{align}
\sin^{2}{\theta}_{13}&\approx \frac{32}{9}\vt\epsilon \vt^{2} \left| \frac{ m^{0}_{\nu_{3}} }{ m^{0}_{\nu_{1}}}\right|^{2}(1-R_{1})^{2},\quad
\sin^{2}{\theta}_{23}\approx  \frac{1}{2}\frac{\left|1+\epsilon \left(1\pm \frac{2}{3} \left|\frac{m^{0}_{\nu_{3}}}{m^{0}_{\nu_{1}}}\right|\right)\right|^{2}}{1-\sin^{2}{\theta_{13}}}.\label{anih}
 \end{align}}
From these formulas, in general, an $\vert \epsilon\vert $ large value will be needed to compensate the $\vert m^{0}_{\nu_{3}}\vert$ lightest neutrino mass to get the allowed region for the reactor angle. But, the atmospheric angle prefers an $\vert \epsilon \vert$ small values. In addition, since that $r_{3}<0$, the complex parameter phase is taken to be $\alpha_{\epsilon}=0$ to increase the atmospheric angle value.
In order to fix ideas, we obtain for the {\bf Case C}: (a) if $\vert \epsilon\vert\approx 0.3$, it is required that $\vert m^{0}_{\nu_{3}}\vert/\vert m^{0}_{\nu_{1}}\vert \approx 0.26$, to obtain $\sin^{2}{\theta_{13}}\approx 0.0229$. As a consequence, we get $\sin^{2}{\theta_{23}}\approx 0.94$ which is too large; (b) if $\vert \epsilon\vert\approx 0.1$, then we need that $\vert m^{0}_{\nu_{3}}\vert/\vert m^{0}_{\nu_{1}}\vert \approx 0.8$ to get $\sin^{2}{\theta_{13}}\approx 0.0229$, and therefore, $\sin^{2}{\theta_{23}}\approx 0.68$, which is still large in comparison to the central value.
\begin{figure}[ht]
\centering
\includegraphics[scale=0.55]{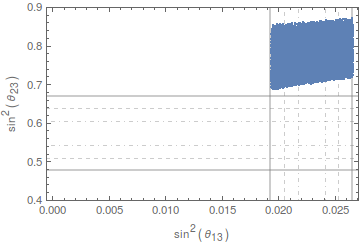}\hspace{0.3cm}\includegraphics[scale=0.55]{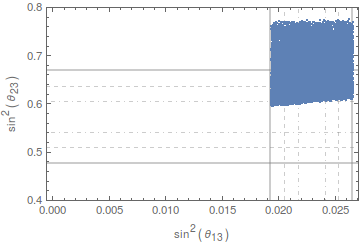}\hspace{0.3cm}
\caption{ $\sin^{2}{\theta_{23}}$ versus $\sin^{2}{\theta_{13}}$. The left and right panels stand for the {\bf Case C}  and {\bf Case D}, respectively. The dotdashed, dashed and thick lines stand for $1~\sigma$, $2~\sigma$ and $3~\sigma$, respectively for each case. \label{fi1}}
\end{figure}

For the {\bf Case D}, the reactor angle has approximately the same values for $\vert \epsilon\vert\approx 0.3$, $\vert \epsilon\vert\approx 0.1$ and their respective $\vert m^{0}_{\nu_{3}}\vert/\vert m^{0}_{\nu_{1}}\vert$ mass ratios as above case. Then, with these values of $\vert \epsilon \vert$, we obtain $\sin^{2}{\theta_{23}}\approx 0.79$ and $\sin^{2}{\theta_{23}}\approx 0.56$, respectively. Notice that both values are approaching the allowed region for this mixing angle, then, this case is more favorable than the {\bf Case C}. This happens since a large contribution of $\vert \epsilon \vert$, in the atmospheric angle, is suppressed by $r_{3}$ which is minor than $1$ and the reactor angle prefers an $\vert \epsilon \vert$ large values.

Let us remark the following, if $\vert \epsilon \vert$ is tiny, we require that $\vert m^{0}_{\nu_{3}}\vert/\vert m^{0}_{\nu_{1}}\vert$ neutrino mass ratio should be larger than $1$ to enhance the reactor angle but this mass ratio violates the inverted ordering. This statement is valid for the {\bf Cases C} and {\bf D}. At the same time, if $\alpha_{\epsilon}=\pi$ is chosen in the atmospheric angle, this would be tiny for the same values of $\vert \epsilon \vert$ and the $\vert m^{0}_{\nu_{3}}\vert/\vert m^{0}_{\nu_{1}}$, as can be verified straight from Eq. (\ref{anih}).

\begin{figure}[ht]
\centering
\includegraphics[scale=0.55]{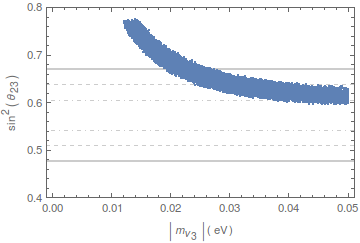}\hspace{0.3cm}\includegraphics[scale=0.55]{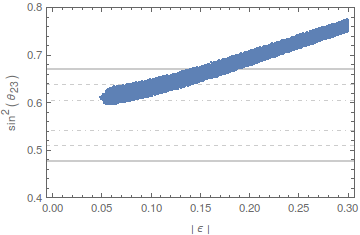}
\caption{{\bf Case D}: Allowed region for $\sin^{2}{\theta_{23}}$. 
The dotdashed, dashed and thick lines stand for $1~\sigma$, $2~\sigma$ and $3~\sigma$ \label{fi2}.}
\end{figure}

In order to get a complete view of the parameter space, let us show some plots for the reactor and atmospheric angles. We have considered the exact formulas given in Eq. (\ref{mixang}), for the the observables as the $\theta_{12}$ solar angle, $\Delta m^{2}_{21}$ and $\Delta m^{2}_{13}$, their values were taken up to $3~\sigma$. Then, the figure {\ref{fi1}} shows the atmospheric angle versus the reactor one for the {\bf Case C} and {\bf D}. This scattering plots clearly support our analytic result on the {\bf Case C}, this is,  both mixing angles can not be accommodate simultaneously. In the {\bf Case D}, the reactor angle is consistent with the experimental data but the atmospheric one is large but consistent up to $2-3~\sigma$ in its allowed region. In addition, for the {\bf Case D}, the parameter space is shown in the figure {\ref{fi2}}. As can be seen, the atmospheric angle prefers small values for $\vert \epsilon \vert$ whereas the reactor one needs a large value, as was already pointed out. Moreover, the set of values for $\vert \epsilon \vert$ and $\vert m^{0}_{\nu_{3}}\vert$ is tight.

{\bf Degenerated hierarchy}. In this case, $\vert m^{0}_{\nu_{3}}\vert \approxeq \vert m^{0}_{\nu_{2}}\vert \approxeq \vert m^{0}_{\nu_{1}}\vert\approxeq m_{0}$, with $m_{0}\gtrsim  0.1~eV$. 
Then, the absolute neutrino masses can be written as $\vert m^{0}_{\nu_{3}} \vert= \sqrt{\Delta m^{2}_{31}+m^{2}_{0}}\approx  m_{0}\left(1+\Delta m^{2}_{31}/2m^{2}_{0}\right)$ and $\vert m^{0}_{\nu_{2}} \vert=\sqrt{\Delta m^{2}_{21}+m^{2}_{0}}\approx m_{0}\left(1+\Delta m^{2}_{21}/2m^{2}_{0}\right)$. As in the inverted case, there are four independent cases for the signs which are shown below.
\begin{itemize}
\item {\bf Case A}. If $m^{0}_{\nu_{i}}>0 $,
{\scriptsize
\begin{align}
r^{A}_{1}=\frac{\vert m^{0}_{\nu_{3}}\vert+m_{0}}{\vert m^{0}_{\nu_{3}}\vert-m_{0}},\qquad r^{A}_{2}=\frac{\vert m^{0}_{\nu_{3}}\vert+\vert m^{0}_{\nu_{2}}\vert}{\vert m^{0}_{\nu_{3}}\vert-\vert m^{0}_{\nu_{2}}\vert},\qquad r^{A}_{3}=r^{A}_{2}\cos^{2}{\theta_{\nu}}+r^{A}_{1}\sin^{2}{\theta_{\nu}}.\label{dh1}
\end{align}}
\item {\bf Case B}. If  If $m^{0}_{\nu_{(2, 1)}}>0$ and $m^{0}_{\nu_{3}}<0$,
{\scriptsize
	\begin{align}
	r^{B}_{1}=\frac{\vert m^{0}_{\nu_{3}}\vert-m_{0}}{\vert m^{0}_{\nu_{3}}\vert+m_{0}}=\frac{1}{r^{A}_{1}},\qquad r^{B}_{2}=\frac{\vert m^{0}_{\nu_{3}}\vert-\vert m^{0}_{\nu_{2}}\vert}{\vert m^{0}_{\nu_{3}}\vert+\vert m^{0}_{\nu_{2}}\vert}=\frac{1}{r^{A}_{2}},\qquad r^{B}_{3}=r^{B}_{2}\cos^{2}{\theta_{\nu}}+r^{B}_{1}\sin^{2}{\theta_{\nu}}.\label{dh4}
	\end{align}}
\item {\bf Case C}. If $m^{0}_{\nu_{(3, 2)}}>0$ and $m^{0}_{\nu_{1}}=-m_{0}$, 
{\scriptsize
\begin{align}
r^{C}_{1}=\frac{\vert m^{0}_{\nu_{3}}\vert-m_{0}}{\vert m^{0}_{\nu_{3}}\vert+m_{0}}=\frac{1}{r^{A}_{1}},\qquad r^{C}_{2}=\frac{\vert m^{0}_{\nu_{3}}\vert+\vert m^{0}_{\nu_{2}}\vert}{\vert m^{0}_{\nu_{3}}\vert-\vert m^{0}_{\nu_{2}}\vert}=r^{A}_{2},\qquad r^{C}_{3}=r^{C}_{2}\cos^{2}{\theta_{\nu}}+r^{C}_{1}\sin^{2}{\theta_{\nu}}.
\label{dh2}
\end{align}}
\item {\bf Case D}. If $m^{0}_{\nu_{2}}>0$ and $m^{0}_{\nu_{(3, 1)}}<0$, 
{\scriptsize
\begin{align}
r^{D}_{1}=\frac{\vert m^{0}_{\nu_{3}}\vert+m_{0}}{\vert m^{0}_{\nu_{3}}\vert-m_{0}}=r^{A}_{1},\qquad r^{D}_{2}=\frac{\vert m^{0}_{\nu_{3}}\vert-\vert m^{0}_{\nu_{2}}\vert}{\vert m^{0}_{\nu_{3}}\vert+\vert m^{0}_{\nu_{2}}\vert}=\frac{1}{r^{A}_{2}}, \qquad r^{D}_{3}=r^{D}_{2}\cos^{2}{\theta_{\nu}}+r^{D}_{1}\sin^{2}{\theta_{\nu}}. \label{dh3}
\end{align}}
\end{itemize}
Notice that 
{\scriptsize
\begin{align}
\vert m^{0}_{\nu_{3}}\vert- m_{0}&\approx 2 m_{0}R_{2},\quad
\vert m^{0}_{\nu_{3}}\vert+ m_{0}\approx 2m_{0}\left(1+R_{2}\right),\nn\\ \vert m^{0}_{\nu_{3}}\vert- \vert m^{0}_{\nu_{2}}\vert &\approx 2m_{0}R_{2}\left(1-R_{3}\right),\quad
\vert m^{0}_{\nu_{3}}\vert+ \vert m^{0}_{\nu_{2}}\vert \approx 2 m_{0}\left[1+R_{2}+R_{4}\right],\label{dh5}
\end{align}}
with $R_{2}\equiv \Delta m^{2}_{31}/4m^{2}_{0}$, $R_{3}=\Delta m^{2}_{21}/\Delta m^{2}_{31}$ and $R_{4}=\Delta m^{2}_{21}/4m^{2}_{0}$,     where $R_{4}< R_{3}\lesssim R_{2}$. 
Thus, $r^{A}_{1}\approx (1+R_{2})/R_{2}$ and $r^{A}_{2}\approx r^{A}_{1}(1+R_{3})$. To fix ideas on the order of magnitude of each defined quantity, we use the data for the inverted hierarchy and their respective central values. So that, $R_{2}\sim 6\times 10^{-2}$,  $R_{3}\sim 3\times 10^{-2}$,  $R_{4}\sim 2\times 10^{-3}$ and $r^{A}_{1}\sim 17$ with $m_{0}=0.1~eV$. Indeed, $R_{2}$ and $R_{4}$ might be fairly small, and therefore $r^{A}_{1}$ so large since that $m_{0}\gtrsim 0.1~eV$.

Therefore, in the {\bf Cases A}, $r^{A}_{2}-r^{A}_{1}\approx r^{A}_{1}R_{3}$ and $r^{A}_{3}\approx r^{A}_{1}$ then
{\scriptsize
\begin{align}
\sin^{2}{\theta_{13}}&\approx \frac{2}{9}\left|\epsilon \right|^{2}\left[r^{A}_{1} R_{3}\right]^{2},\qquad \sin^{2}{\theta_{23}}\approx \frac{1}{2}\frac{\left|1-\epsilon r^{A}_{1}\right|^{2}}{1-\sin^{2}{\theta_{13}}}.
\end{align}}
In the {\bf Case B}, $r^{B}_{2}-r^{B}_{1}\approx -r^{A}_{1}/R_{3}$ and $r^{B}_{3}\approx 1/r^{A}_{1}$ so that
{\scriptsize
\begin{align}
\sin^{2}{\theta_{13}}&\approx \frac{2}{9}\left|\epsilon \right|^{2}\left[ \frac{R_{3}}{r^{A}_{1}}\right]^{2},\qquad \sin^{2}{\theta_{23}}\approx \frac{1}{2}\frac{\left|1-\frac{\epsilon}{ r^{A}_{1}}\right|^{2}}{1-\sin^{2}{\theta_{13}}}.
\end{align}	
}
Thus, in the former case if the reactor angle is fixed to its central value ($\sin^{2}{\theta_{13}}\approx 0.0229$), with the above values for $r^{A}_{1}$ and $R_{3}$, we obtain that $\vert \epsilon \vert \approx 0.5$ which means a strong breaking of the $\mu-\tau$
symmetry. As result, the atmospheric angle comes out begin too large. Analogously, for the second case one gets that $\vert \epsilon \vert\approx 10^{2}$ if the reactor angle is fixed to its central value. As consequence, the atmospheric angle is also quite large. Therefore, these two cases are excluded, the only feasible cases are the last ones.

For the {\bf Case C}, from Eq.(\ref{dh2}), we have $r^{C}_{2}-r^{C}_{1}\approx r^{A}_{1}(1+R_{3})$ and $r^{C}_{3}\approx r^{A}_{1}\cos^{2}{\theta_{\nu}}$, so that
{\scriptsize
\begin{align}
\sin^{2}{\theta}_{13}\approx\frac{2}{9}\vert \epsilon \vert^{2}\left[r^{A}_{1}(1+R_{3})\right]^{2},\qquad \sin^{2}{\theta_{23}}\approx \frac{1}{2}\frac{\left|  1-\frac{2}{3}r^{A}_{1}\epsilon \right|^{2} }{1-\sin^{2}{\theta_{13}}}.
\label{dh8}
\end{align}}

For the {\bf Case D}, from Eq. (\ref{dh3}), we obtain $r^{D}_{2}-r^{D}_{1}\approx -r^{A}_{1}$ and $r^{D}_{3}\approx r^{A}_{1}\sin^{2}{\theta_{\nu}}$. Then,
{\scriptsize
\begin{align}
\sin^{2}{\theta}_{13}\approx \frac{2}{9}\vert \epsilon \vert^{2} \left[r^{A}_{1} \right]^{2},\qquad \sin^{2}{\theta_{23}}\approx \frac{1}{2}\frac{\left| 1-\frac{1}{3}r^{A}_{1}\epsilon  \right|^{2}}{1-\sin^{2}{\theta_{13}}}.\label{dh9}
\end{align}}
\begin{figure}[ht]
	\centering
	\includegraphics[scale=0.55]{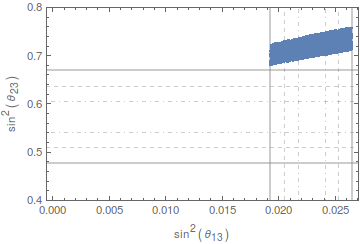}\hspace{0.3cm}\includegraphics[scale=0.55]{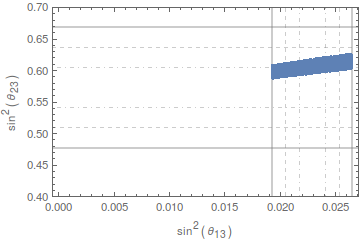}
	\caption{$\sin^{2}{\theta_{23}}$ versus $\sin^{2}{\theta_{13}}$. The left and right panels stand for the {\bf Case C}  and {\bf Case D}, respectively. The dotdashed, dashed and tick lines stand for $1~\sigma$, $2~\sigma$ and $3~\sigma$, respectively for each case. \label{fd1}}
\end{figure}
Roughly speaking, as in the inverted case, the reactor angle has approximately the same behavior for both cases but the atmospheric angle comes out being different. In here, on the other hand, notice that $r_{3}>0$ for both cases then if $\alpha_{\epsilon}=0$, the atmospheric angle would be smaller than $45^{\circ}$ which is far away from the experimental data, as can be verified from Eq.(\ref{dh8}) and Eq.(\ref{dh9}). In order to increase this value, it is needed that $\alpha_{\epsilon}=\pi$. Additionally, because of $r^{A}_{1}\ggg 1$, the value of $\vert \epsilon \vert$ should be of the order of $10^{-2}$ in order to not enhance too much the atmospheric angle, of course, we must be careful to not spoil the reactor angle or vice versa. 

Now, in {\bf Case C} and {\bf D}, if the reactor angle is fixed to its central value ($\sin^{2}{\theta_{13}}\approx 0.0229$) then it is required that $\vert \epsilon \vert\sim 2\times 10^{-2}$, so that one obtains $\sin^{2}{\theta_{23}}\approx 0.74$ and $\sin^{2}{\theta_{23}}\approx 0.63$, respectively. As can be seen, the favored case is the latter due to the $\epsilon r^{D}_{3}$ contribution, in the atmospheric angle, is minor than $\epsilon r^{C}_{3}$ such that the atmospheric angle is softly being deviated from $45^{\circ}$. Now, an interesting fact is the following: if $m_{0}$ is increased to the allowed value, then $r^{A}_{1}$ becomes quite large and therefore, a tiny  $\vert \epsilon \vert$ value is needed to not deviate so much from $45^{\circ}$ the atmospheric angle and at the same time, to get an allowed region for the reactor angle. In this hierarchy, the $\mu-\tau$ symmetry is being broken softly.
	
\begin{figure}[ht]
\centering
\includegraphics[scale=0.55]{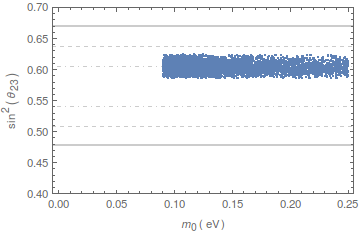}\hspace{0.3cm}\includegraphics[scale=0.55]{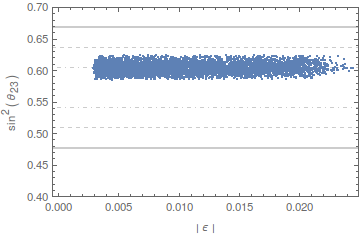}
\caption{{\bf Case D}: Allowed region for $\sin^{2}{\theta_{23}}$. 
The dotdashed, dashed and thick lines stand for $1~\sigma$, $2~\sigma$ and $3~\sigma$\label{fd2}}
\end{figure}

We will now explore the complete parameter space for both cases. The exact formulas for the mixing angles have been used with the respective extreme Majorana phases for each case, apart from the allowed values for $\Delta m^{2}_{21}$, $\Delta m^{2}_{13}$ and $\theta_{\nu}$ the solar angle for the inverted ordering as a good approximation. Therefore, in figure \ref{fd1}, the atmospheric versus the reactor angle is show up to $3~\sigma$. This panels allow to compare the two cases and these support our analytic result, in the {\bf Case D}, both angles of interest are accommodated very well. In the figure \ref{fd2}, as can be seen, the parameter space is large where  the atmospheric angle, and therefore the reactor one, is accommodated in good agreement with the experimental data. At the end of the day, the degenerate ordering is favored instead of the inverted case. 

\section{Conclusions}
We have extended the scalar sector of the LRSM in order to get masses and mixings for fermions. In the lepton sector, neutrino masses and mixings have been studied in the limit of a slightly broken $\mu-\tau$ symmetry, so that the reactor and atmospheric angles depend strongly on the $\epsilon$ free parameter, that characterizes the $\mu-\tau$ symmetry breaking, and the neutrino masses. Due to this last fact, the mixing angles are sensitive to the extreme Majorana phases which may increase or decrease their respective values. Therefore, we have made an analytic study on the role that the extreme Majorana phases might have in each hierarchy. Additionally, the $\epsilon$ free parameter and the lightest neutrino mass have been constrained.

The main results are the following: (a) the model predicts a tiny value for the reactor angle in the normal hierarchy and this result holds for whatever extreme Majorana phases. Then, the normal ordering is completely ruled out for $\vert \epsilon\vert \leq 0.3$; (b) in the inverted hierarchy
there is one combination in the extreme Majorana phases where the reactor and atmospheric angles are compatible up to $2-3~\sigma$ within the allowed region for the latter angle. This scenario is fairly constrained since the parameter space is so tight; (c) the degenerate ordering is the most viable scenario to accommodate simultaneously the reactor and atmospheric angles. In this case, there is one combination in the extreme Majorana phases where both angles are consistent with the current limits imposed  by the experimental data for $\sin^{2}\theta_{23}$ and $\sin^{2}\theta_{13}$.  At the same time, a set of values for $\epsilon$ and the lightest neutrino mass was found such that the $\mu-\tau$ symmetry is broken softly. Remarkably, the viable cases predict that $\theta_{23}>45^{\circ}$.

For the moment, the quark sector has been left aside for a future work but we have pointed out that the mass matrices possess textures that might fit the CKM matrix. Although the model is quite elaborate, it is fairly  predictive and testable by the future results that the Nova and KamLAND-Zen collaborations will provide.

\section*{Acknowledgements}
We would like to thank Myriam Mondrag\'on and Abdel P\'erez-Lorenzana for their useful comments and discussion on the manuscript. This work was partially supported by a PAPIIT grant IN111115. The author thanks Red de Altas Energ\'{\i}as-CONACYT for the financial support.
   
\bibliographystyle{unsrt}
\bibliography{references.bib}
\end{document}